\documentclass{article}

\usepackage{amsmath,amsfonts,amssymb}
\usepackage{color}
\usepackage{bm}
\usepackage{graphicx}
\usepackage{a4wide}

\title{Dynamical supersymmetry on the XXX spin chain}
\author{Chihiro Matsui \\[3ex]
{\it Department of Mathematical Informatics, The University of Tokyo} \\
{\it 7-3-1 Hongo, Bunkyo-ku, Tokyo 113-8656, Japan}}

\begin{document}
\maketitle

\abstract{
We show the XXX model has the $\mathcal{N}=2$ dynamical supersymmetry. Using the supercharges defined by the Jordan-Wigner fermions, it was found that the anti-commutation relation of the supercharges gives the Hamiltonian of the XXX model with magnetic field. In order to compare the length-change supercharges with the conventional ones, we interpreted their actions in the spinon basis. In the last part of this paper, we propose the application of the dynamical supersymmetry to the models with magnetic impurities through the Bethe-ansatz analysis. 
}

\section{Introduction}
It has been an interesting problem why an effective field theory of a lattice model posses the symmetry which is not explicitly obtained in a lattice model. For instance, both the spin-$1$ XXZ model and its effective field theory, {\it i.e.} the supersymmetric sine-Gordon model, have the $U_q(sl_2)$-symmetry in their scattering process~\cite{bib:ZF80, bib:B82, bib:B83}, worthwhile only the latter has the supersymmetry~\cite{bib:IO93}. Indeed, to define the supersymmetry on a lattice itself has been a longstanding problem, as the Poincare symmetry is broken on a lattice, and therefore usually the extension of the Poincare symmetry is required. 
One way to resolve this problem was provided by Yang and Fendley~\cite{bib:YF04}. In order to define the supersymmetry on spin chains, they introduced supercharges which change the length of chains by one. They found the supersymmetry on the spin-$1/2$ XYZ chain for the special anisotropy, and also for the higher-spin cases. Especially, in the spin-$1/2$ XXZ case, the supersymmetry emerges at the Razumov-Stroganov point~\cite{bib:RS04}. 

The length-change model plays an important role also in the context of the AdS/CFT correspondence. Up to the one-loop correction, the scaling dimensions of gauge invariant operators in the $\mathcal{N} = 4$ super Yang-Mills (SYM) theory is computed from the XXX spin chain~\cite{bib:MZ03, bib:BS03}. However, if we further compute higher-loop corrections, it is inevitable to encounter length-change of the XXX spin chain~\cite{bib:BKS03, bib:B04}. The idea of dynamic spin chains has been introduced to the context of the $\mathcal{N} = 4$ superconformal gauge theory by Beisert~\cite{bib:B04} and he found that this dynamical spin chain is still integrable. Thus, studying the supersymmetry on a dynamical spin chain gives new insights not only to the new symmetry on the spin chain but also to computation of higher-loop correction of the $\mathcal{N} = 4$ SYM theory. 

In this paper, we show that the isotropic (XXX) model also posses the supersymmetry. Unlike the Yang-Fendley supercharges, which are defined by fermions of electron orbitals, we define supercharges by the Jordan-Wigner fermions. We found that the anti-commutation relation of our supercharges results in the Hamiltonian of the XXX spin chain with magnetic field. The existence of superpartners is predicted from the Bethe-ansatz analysis. 
Since the conventional supersymmetry defined on a field theory has been discussed in the particle basis, we discuss our supercharge in the spinon basis, which becomes the particle basis in the continuum limit. The supercharge acts as a creation or annihilation operator of a zero-momentum spinon and thus, the actions of supercharges are interpreted just as exchanging the domains which spinons are carrying. 

This paper is organized as follows. In the next section, we give the definition of supersymmetry. At the same time, we show what motivates us to consider the supersymmetry on the XXX spin chain. In the section 3, we introduce the supercharges in terms of the Jordan-Wigner fermions. These newly-introduced supercharges are checked to satisfy the super algebra and it is also showed that their anti-commutation relation gives the Hamiltonian of the XXX model with magnetic field. In the section 4, the supersymmetry in the spinon basis is discussed. The actions of the supercharges are investigated on the $n_s$-particle spinon state. The result verifies the Bethe-ansatz analysis about superpartners. The last section is devoted to the concluding remarks, in which the application to the models with magnetic impurities are proposed.

\section{Defining relations of the supersymmetry}
The $\mathcal{N}=2$ supersymmetry is defined by the nilpotency of the two supercharges $Q^2 = (Q^{\dag})^2 = 0$ and the anti-commutation relation:
\begin{equation} \label{anti-commutation}
 \{Q,\, Q^{\dag}\} = H, 
\end{equation}
which gives the Hamiltonian. The additional commutation relations are satisfied with the fermion number operator $F$:
\begin{equation} \label{commutation}
 [F,\, Q] = -Q, 
  \qquad
  [F, Q^{\dag}] = Q^{\dag}. 
\end{equation}
The supercharges exchange fermions and bosons, {\it i.e.} if we write a fermion state by $|f\rangle$ and a boson state by $|b\rangle$, we have 
\begin{equation}\label{fermion-boson}
 Q|f\rangle = \sqrt{E}|b\rangle, 
  \qquad
  Q|b\rangle = \sqrt{E}|f\rangle. 
\end{equation}
The fermion number operator distinguishes a fermion state and a boson state: 
\begin{equation}
 (-1)^F |f\rangle = -|f\rangle,
  \qquad
  (-1)^F |b\rangle = |b\rangle. 
\end{equation}
The important fact is that, in the non-zero energy case $E \neq 0$, the two states $|f\rangle$ and $|b\rangle$ connected by the supercharge make a doublet called superpartners~\cite{bib:W82}. 

Recently, Yang and Fendley introduced the dynamical supersymmetry on the XXZ spin chain~\cite{bib:YF04} which adds or removes a lattice site. Their idea is based on the transformation \eqref{fermion-boson}, which removes a fermion, if the supercharge is applied on a fermion state. In the first paper~\cite{bib:YF04} of the series of their papers, they have found that the supersymmetry emerges in the spin-$1/2$ XXZ chain at the Razumov-Stroganov point where the anisotropy is given by $\Delta = -1/2$. Later, they extended their method to define the dynamical supersymmetry to the XYZ case~\cite{bib:HF12} and the higher-spin cases~\cite{bib:H13}. At the same time, they gave the Bethe-ansatz viewpoint of the superpartners. Transforming the Bethe equations for the length-$N$ spin-$1/2$ XXZ spin chain with $n$ roots: 
\begin{equation} \label{be}
 \left(\frac{\sinh(\lambda_j + \frac{i\gamma}{2})}{\sinh(\lambda_j - \frac{i\gamma}{2})}\right)^N
  =
  \prod_{k=1 \atop k \neq j}^n \frac{\sinh(\lambda_j - \lambda_k + i\gamma)}{\sinh(\lambda_j - \lambda_k - i\gamma)}, 
\end{equation}
where $\gamma$ is the anisotropy parameter given by $\gamma = \arccos \Delta$, they defined the Bethe equations for the one-less-site with one-more-root: 
\begin{equation} \label{be_fendley}
 \left(\frac{\sinh(\lambda_j + \frac{i\gamma}{2})}{\sinh(\lambda_j - \frac{i\gamma}{2})}\right)^{N-1}
  =
  \prod_{k=1 \atop k \neq j}^n \frac{\sinh(\lambda_j - \lambda_k + i\gamma)}{\sinh(\lambda_j - \lambda_k - i\gamma)}
  \frac{\sinh(\lambda_j - \frac{i\gamma}{2})}{\sinh(\lambda_j + \frac{i\gamma}{2})},
\end{equation}
by regarding the last factor of the right-hand side as the $(n+1)$th root. This identification is allowed only when $\gamma = \pi/3$ and the $(n+1)$th root is given by $\lambda_{n+1} = i\pi$. 
The transformation \eqref{be_fendley} implies that the Bethe state on the length-$N$ chain given by the roots $\lambda_1,\dots,\lambda_n$ makes the superpartner with the Bethe state of the length-$(N-1)$ chain given by the same roots besides $\lambda_{n+1} = i\pi$. 
Thus, the Yang-Fendley supercharges transform a down spin into two up spins, or vice versa. 

The Bethe equations \eqref{be} admit another transformation to describe the system one-site-more together with the extra $(n+1)$th root: 
\begin{equation} \label{be_matsui}
 \left(\frac{\sinh(\lambda_j + \frac{i\gamma}{2})}{\sinh(\lambda_j - \frac{i\gamma}{2})}\right)^{N+1}
  =
  \prod_{k=1 \atop k \neq j}^n \frac{\sinh(\lambda_j - \lambda_k + i\gamma)}{\sinh(\lambda_j - \lambda_k - i\gamma)}
  \frac{\sinh(\lambda_j + \frac{i\gamma}{2})}{\sinh(\lambda_j - \frac{i\gamma}{2})}. 
\end{equation}
This transformation is allowed when $\gamma = 2\pi$ and the extra root is given by $\lambda_{n+1} = i\pi$, that is, the $N$-site Bethe state with the roots $\lambda_1,\dots,\lambda_n$ has the same energy as the $(N+1)$-site Bethe state given by the same roots together with $\lambda_{n+1} = i\pi$. This motivates us to find a dynamical supersymmetry not only at the $\gamma = \pi/3$ case but also $\gamma = 2\pi$, {\it i.e.} the XXX point. 

\section{Supercharges as Jordan-Wigner fermions}
Now we introduce the supercharges. Our supercharges are defined in terms of the Jordan-Wigner fermions, instead of electron-orbital fermions used in Yang-Fendley supercharges. First we define the local supercharges in such a way that increase (resp. decrease) the number of lattice site by one at the same time with adding (resp. removing) a down spin. By keeping the position of the $j$th site, there are two ways of the actions of supercharges, that is, either 
\begin{equation}
 Q_j^{\dag}:\;|\dots\,\underset{j}{\uparrow}\,\dots\rangle \mapsto |\dots\,\underset{j+\frac{1}{2}}{\uparrow} \underset{j-\frac{1}{2}}{\downarrow}\,\dots \rangle, 
\end{equation} 
or 
\begin{equation}
 Q_j^{\dag}:\;|\dots\,\underset{j}{\uparrow}\,\dots\rangle \mapsto |\dots\,\underset{j-\frac{1}{2}}{\downarrow} \underset{j+\frac{1}{2}}{\uparrow}\,\dots \rangle, 
\end{equation}
and their conjugate actions. We call the former transformation the left action, since the up arrow shifts to the right by a half site, while the latter transformation the right action. In order to deal with the left and right actions separately, we write the supercharges in the following matrix forms: 
\begin{equation} \label{localsupercharge_matsui} 
  Q_j := 
   \begin{pmatrix}
    R_{j-\frac{1}{2}} \psi_{j-\frac{1}{2}} & 0 \\ 
    0 & R_{j+\frac{1}{2}} \psi_{j+\frac{1}{2}}
   \end{pmatrix}, 
   \qquad
 Q_j^{\dag} := 
 \begin{pmatrix}
  \psi_{j-\frac{1}{2}}^{\dag} R_{j-\frac{1}{2}}^{\dag} & 0 \\ 
  0 & \psi_{j+\frac{1}{2}}^{\dag} R_{j+\frac{1}{2}}^{\dag}
 \end{pmatrix},
\end{equation}
where $R_j^{\dag}$ shifts all the sites after $j$th site to the right by a half site and the other sites to the left by a half site by leaving the $(j-1/2)$th site unoccupied. Thus, after applying the supercharges, the lattice sites locate at the middle of the original location of sites. Without loss of generality, we set lattice fermions on half-integer sites for an even-length chain, while on integer sites for an odd-length chain. Therefore, $j$ takes either a half-integer or integer value depending on the length of the spin chain. 
In order to define supercharges on the periodic chain, we define the edge operators $Q_{\frac{1}{2}}$ and $Q_N^{\dag}$ which act as 
\begin{equation}
 Q_{\frac{1}{2}} := 
   \begin{pmatrix}
    R_N \psi_N & 0 \\
    0 &R_1 \psi_1
   \end{pmatrix}, 
   \qquad
   Q_N^{\dag} := 
   \begin{pmatrix}
    \psi_{N-\frac{1}{2}}^{\dag} R_{N-\frac{1}{2}}^{\dag} & 0 \\ 
    0 & \psi_{\frac{1}{2}}^{\dag} R_{\frac{1}{2}}^{\dag} 
   \end{pmatrix}
\end{equation}
%For convenience, we consider that lattice sites are integer for the even-length chain and half-integer for the odd-length chain. 
for an odd-length chain. Note that the periodicity requires the $0$th site to be identified with the $(N+1)$th site. The edge actions of supercharges for an even-length chain are similarly defined. 

From now on, we only discuss the odd $N$ case. The even $N$ case is discussed just by replacing integer $j$ by half-integer $j$. The supercharges are defined by the summation of local supercharges: 
\begin{equation} \label{supercharge_matsui}
 Q = \sum_{j=1}^N Q_j, 
  \qquad
  Q^{\dag} = \sum_{j=1}^{N} Q_{j-\frac{1}{2}}^{\dag}. 
\end{equation}
Besides, we introduce the fermion number operator defined by 
\begin{equation}
 F = \sum_{j=1}^N F_j := \sum_{j=1}^N \psi_j^{\dag} \psi_j, 
\end{equation}
which just counts the number of down spins. Corresponding to the left and right actions , we introduce the left state $|\Psi_{\rm L} \rangle$ and the right state $|\Psi_{\rm R} \rangle$. The supercharges act on the left and right states as  
\begin{align}
 &Q |\bm{\Psi} \rangle 
 = \sum_{j=1}^N
 \begin{pmatrix}
  R_j \psi_j & 0 \\ 
  0 & R_{j+1} \psi_{j+1}
 \end{pmatrix}
 |\bm{\Psi} \rangle, 
 \\
 &Q^{\dag} |\bm{\Psi} \rangle 
 = \sum_{j=1}^N
 \begin{pmatrix}
  \psi_{j-\frac{1}{2}}^{\dag} R_{j-\frac{1}{2}}^{\dag} & 0 \\ 
  0 & \psi_{j+\frac{1}{2}}^{\dag} R_{j+\frac{1}{2}}^{\dag}
 \end{pmatrix}
 |\bm{\Psi} \rangle, 
\end{align}
where $|\bm{\Psi} \rangle := (\Psi_{\rm L}\, \Psi_{\rm R})^{\mathsf{T}}$

The non-zero actions of the supercharges \eqref{supercharge_matsui} are explicitly written as  
\begin{align}
 &Q_j 
 \begin{pmatrix}
  |\dots\,\underset{j-\frac{1}{2}}{\uparrow} \underset{j+\frac{1}{2}}{\downarrow}\,\dots\rangle \\
  |\dots\,\underset{j-\frac{1}{2}}{\downarrow} \underset{j+\frac{1}{2}}{\uparrow}\,\dots\rangle
 \end{pmatrix} 
 = (-1)^{F_1} \dots (-1)^{F_{j-1}} q_j 
 \begin{pmatrix}
  |\dots\,\underset{j-\frac{1}{2}}{\uparrow} \underset{j+\frac{1}{2}}{\downarrow}\,\dots\rangle \\
  |\dots\,\underset{j-\frac{1}{2}}{\downarrow} \underset{j+\frac{1}{2}}{\uparrow}\,\dots\rangle 
 \end{pmatrix} 
 = (-1)^{m_j} 
 \begin{pmatrix}
  |\dots\,\underset{j}{\uparrow}\,\dots\rangle \\
  |\dots\,\underset{j}{\uparrow}\,\dots\rangle
 \end{pmatrix}, 
 \\
 &Q_j^{\dag} 
 \begin{pmatrix}
  |\dots\,\underset{j}{\uparrow}\,\dots\rangle \\
  |\dots\,\underset{j}{\uparrow}\,\dots\rangle
 \end{pmatrix}
 = (-1)^{F_1} \dots (-1)^{F_{j-1}} q_j^{\dag} 
 \begin{pmatrix}
  |\dots\,\underset{j}{\uparrow}\,\dots\rangle \\
  |\dots\,\underset{j}{\uparrow}\,\dots\rangle
 \end{pmatrix} 
 = (-1)^{m_j} 
 \begin{pmatrix}
  |\dots\,\underset{j-\frac{1}{2}}{\uparrow} \underset{j+\frac{1}{2}}{\downarrow}\,\dots\rangle \\
  |\dots\,\underset{j-\frac{1}{2}}{\downarrow} \underset{j+\frac{1}{2}}{\uparrow}\,\dots\rangle
 \end{pmatrix}, \\
 &Q_N 
 \begin{pmatrix}
  |\underset{\frac{1}{2}}{\downarrow}\,\dots\,\underset{N-\frac{1}{2}}{\uparrow}\rangle \\
  |\underset{\frac{1}{2}}{\uparrow}\,\dots\,\underset{N-\frac{1}{2}}{\downarrow}\rangle 
 \end{pmatrix} 
 = (-1)^{F_1} \dots (-1)^{F_{N-1}} q_0 
 \begin{pmatrix}
  |\underset{\frac{1}{2}}{\downarrow}\,\dots\,\underset{N-\frac{1}{2}}{\uparrow}\rangle \\
  |\underset{\frac{1}{2}}{\uparrow}\,\dots\,\underset{N-\frac{1}{2}}{\downarrow}\rangle 
 \end{pmatrix}
 = (-1)^{m_{N-1}} 
 \begin{pmatrix}
  |\dots\,\underset{N}{\uparrow}\rangle \\
  |\dots\,\underset{N}{\uparrow}\rangle
 \end{pmatrix}, 
 \\
 &Q_N^{\dag} 
 \begin{pmatrix}
  |\dots\,\underset{N}{\uparrow}\rangle \\
  |\dots\,\underset{N}{\uparrow}\rangle
 \end{pmatrix}
 = (-1)^{F_1} \dots (-1)^{F_{N-1}} q_N 
 \begin{pmatrix}
  |\dots\,\underset{N}{\uparrow}\rangle \\
  |\dots\,\underset{N}{\uparrow}\rangle 
 \end{pmatrix}
 = (-1)^{m_{N-1}} 
 \begin{pmatrix}
  |\underset{\frac{1}{2}}{\downarrow}\,\dots\,\underset{N-\frac{1}{2}}{\uparrow}\rangle \\
  |\underset{\frac{1}{2}}{\uparrow}\,\dots\,\underset{N-\frac{1}{2}}{\downarrow}\rangle
 \end{pmatrix}, 
\end{align}
where the operator $q_j$ just transform an up spin at the $j$th site to a set of an up spin and a down spin located at $j-\frac{1}{2}$ and $j+\frac{1}{2}$. The actions are compatible with the conventional supercharges~\cite{bib:S90} whose spacial extensions are defined through the comultiplication~\cite{bib:S90}: 
\begin{align}
 &\Delta(Q) = Q \otimes 1 + (-1)^F \otimes Q, \qquad
  \Delta(Q^{\dag}) = Q^{\dag} \otimes 1 + (-1)^F \otimes Q^{\dag}, \\
 &\Delta(F) = F \otimes 1 + 1 \otimes F. 
\end{align}
For simplifying the notation, we introduced $m_j$ which counts the number of down spins before the $j$th site on the original chain. 

Now we check whether the operators given by \eqref{supercharge_matsui} satisfy the superalgebra. The nilpotency of the supercharges are immediately verified, since $Q$ acts on integer sites by transforming them into half-integer sites, while $Q^{\dag}$ acts on half-integer sites by transforming them into integer sites. Therefore, we starts from checking the anti-commutation relation \eqref{anti-commutation}. Here we introduce the following deformed anti-commutation relation:
\begin{equation}
 \{Q,\,Q^{\dag}\} = \sum_{i,j=1}^N (Q_i Q_j^{\dag} + Q_{i-\frac{1}{2}}^{\dag} Q_{j-\frac{1}{2}}).  
\end{equation}
In the $\mathcal{N}=2$ supersymmetry, the anti-commutation relation of supercharges gives the Hamiltonian~\cite{bib:W82}. Due to the existence of the left and right states, our Hamiltonian is given by 
\begin{equation}
 H = \langle \bm{\Psi}| \{Q,\,Q^{\dag}\} |\bm{\Psi} \rangle. 
\end{equation}
The anti-commutation relation consists of the following distinguishing quadratic operations: 
\begin{equation} \label{hamiltonian}
\begin{split}
 \{Q,\,Q^{\dag}\} 
 &= \sum_{i<j} (Q_{j+1} Q_i^{\dag} + Q_{i+\frac{1}{2}}^{\dag} Q_{j+\frac{1}{2}})
 + \sum_{i>j} (Q_j Q_{i+1}^{\dag} + Q_{i+\frac{1}{2}}^{\dag} Q_{j+\frac{1}{2}}) \\
 &+ \sum_{i} (Q_{i+1} Q_i^{\dag} + Q_i Q_{i+1}^{\dag} 
 + Q_i Q_i^{\dag} + Q_{i+\frac{1}{2}}^{\dag} Q_{i+\frac{1}{2}}). 
\end{split}
\end{equation}
The first term cancels each other out, since each of the first term acts on the same state with different signs: 
\begin{align}
 &Q_{j+1} Q_i^{\dag} 
 \begin{pmatrix}
  |\dots\,\underset{i}{\uparrow}\,\dots\,\underset{j}{\downarrow} \underset{j+1}{\uparrow}\,\dots\rangle \\
  |\dots\,\underset{i}{\uparrow}\,\dots\,\underset{j}{\uparrow}\ \underset{j+1}{\downarrow}\,\dots\rangle
 \end{pmatrix}
 =
 (-1)^{m_i + m_j + 1} 
 \begin{pmatrix}
  |\dots\,\underset{i}{\uparrow} \underset{i+1}{\downarrow}\,\dots\,\underset{j+1}{\uparrow}\,\dots\rangle \\
  |\dots\,\underset{i}{\downarrow} \underset{i+1}{\uparrow}\,\dots\,\underset{j+1}{\uparrow}\,\dots\rangle
 \end{pmatrix}, \\
 &Q_{i+\frac{1}{2}}^{\dag} Q_{j+\frac{1}{2}} 
 \begin{pmatrix}
  |\dots\,\underset{i}{\uparrow}\,\dots\,\underset{j}{\downarrow} \underset{j+1}{\uparrow}\,\dots\rangle \\
  |\dots\,\underset{i}{\uparrow}\,\dots\,\underset{j}{\uparrow} \underset{j+1}{\downarrow}\,\dots\rangle
 \end{pmatrix}
 =
 (-1)^{m_i + m_j} 
 \begin{pmatrix}
  |\dots\,\underset{i}{\uparrow} \underset{i+1}{\downarrow}\,\dots\,\underset{j+1}{\uparrow}\,\dots\rangle \\
  |\dots\,\underset{i}{\downarrow} \underset{i+1}{\uparrow}\,\dots\,\underset{j+1}{\uparrow}\,\dots\rangle
 \end{pmatrix}. 
\end{align}
Similarly, the second term also cancels each other out: 
\begin{align}
 &Q_j Q_{i+1}^{\dag} 
 \begin{pmatrix}
  |\dots\,\underset{j}{\uparrow} \underset{j+1}{\downarrow}\,\dots\,\underset{i+1}{\uparrow}\,\dots\rangle \\
  |\dots\,\underset{j}{\downarrow} \underset{j+1}{\uparrow}\,\dots\,\underset{i+1}{\uparrow}\,\dots\rangle 
 \end{pmatrix}
 =
 (-1)^{m_i + m_j} 
 \begin{pmatrix}
  |\dots\,\underset{j}{\uparrow}\,\dots\,\underset{i}{\uparrow} \underset{i+1}{\downarrow}\,\dots\rangle \\
  |\dots\,\underset{j}{\uparrow}\,\dots\,\underset{i}{\downarrow} \underset{i+1}{\uparrow}\,\dots\rangle
 \end{pmatrix}, \\ 
 &Q_{i+\frac{1}{2}}^{\dag} Q_{j+\frac{1}{2}} 
 \begin{pmatrix}
  |\dots\,\underset{j}{\uparrow} \underset{j+1}{\downarrow}\,\dots\,\underset{i+1}{\uparrow}\,\dots\rangle \\
  |\dots\,\underset{j}{\downarrow} \underset{j+1}{\uparrow}\,\dots\,\underset{i+1}{\uparrow}\,\dots\rangle 
 \end{pmatrix}
 =
 (-1)^{m_i + m_j - 1} 
 \begin{pmatrix}
  |\dots\,\underset{j}{\uparrow}\,\dots\,\underset{i}{\uparrow} \underset{i+1}{\downarrow}\,\dots\rangle \\
  |\dots\,\underset{j}{\uparrow}\,\dots\,\underset{i}{\downarrow} \underset{i+1}{\uparrow}\,\dots\rangle
 \end{pmatrix}. 
\end{align}
The third term acts as the spin-exchange terms: 
\begin{align}
 &Q_i Q_{i+1}^{\dag} 
 \begin{pmatrix}
  |\dots\,\underset{i}{\downarrow} \underset{i+1}{\uparrow}\,\dots\rangle \\
  |\dots\,\underset{i}{\downarrow} \underset{i+1}{\uparrow}\,\dots\rangle
 \end{pmatrix}
 = (-1)^{m_i + 1} Q_i 
 \begin{pmatrix}
  |\dots\,\underset{i-\frac{1}{2}}{\downarrow} \underset{i+\frac{1}{2}}{\uparrow} \underset{i+\frac{3}{2}}{\downarrow}\,\dots\rangle \\
  |\dots\,\underset{i-\frac{1}{2}}{\downarrow} \underset{i+\frac{1}{2}}{\downarrow} \underset{i+\frac{3}{2}}{\uparrow}\,\dots\rangle 
 \end{pmatrix}
 = (-1)^{2m_i + 1} 
 \begin{pmatrix}
  |\dots\,\underset{i}{\uparrow} \underset{i+1}{\downarrow}\,\dots\rangle \\
  0
 \end{pmatrix}, \\
 &Q_{i+1} Q_i^{\dag} 
 \begin{pmatrix}
  |\dots\,\underset{i}{\uparrow} \underset{i+1}{\downarrow}\,\dots\rangle \\
  |\dots\,\underset{i}{\uparrow} \underset{i+1}{\downarrow}\,\dots\rangle
 \end{pmatrix}
 = (-1)^{m_i} Q_{i+1} 
 \begin{pmatrix}
  |\dots\,\underset{i-\frac{1}{2}}{\uparrow} \underset{i+\frac{1}{2}}{\downarrow} \underset{i+\frac{3}{2}}{\downarrow}\,\dots\rangle \\
  |\dots\,\underset{i-\frac{1}{2}}{\downarrow} \underset{i+\frac{1}{2}}{\uparrow} \underset{i+\frac{3}{2}}{\downarrow}\,\dots\rangle 
 \end{pmatrix}
 = (-1)^{2m_i + 1} 
 \begin{pmatrix}
  0 \\
  |\dots\,\underset{i}{\downarrow} \underset{i+1}{\uparrow}\,\dots\rangle
 \end{pmatrix}. 
\end{align}
The last term counts specific spin-configurations. Each of the last term acts as 
\begin{align}
 &Q_i Q_i^{\dag} 
 \begin{pmatrix}
  |\dots\,\underset{i}{\uparrow}\,\dots\rangle \\
  |\dots\,\underset{i}{\uparrow}\,\dots\rangle
 \end{pmatrix}
 = (-1)^{2m_i} 
 \begin{pmatrix}
  |\dots\,\underset{i}{\uparrow}\,\dots\rangle \\
  |\dots\,\underset{i}{\uparrow}\,\dots\rangle
 \end{pmatrix}, \\ 
 &Q_{i+\frac{1}{2}}^{\dag} Q_{i+\frac{1}{2}} 
 \begin{pmatrix}
  |\dots\,\underset{i}{\uparrow} \underset{i+1}{\downarrow}\,\dots\rangle \\
  |\dots\,\underset{i}{\downarrow} \underset{i+1}{\uparrow}\,\dots \rangle 
 \end{pmatrix}
 = (-1)^{2m_i} 
 \begin{pmatrix}
  |\dots\,\underset{i}{\uparrow} \underset{i+1}{\downarrow}\,\dots\rangle \\
  |\dots\,\underset{i}{\downarrow} \underset{i+1}{\uparrow}\,\dots\rangle
 \end{pmatrix}. 
\end{align}
Thus, the Hamiltonian \eqref{hamiltonian} is written in terms of the spin operators as 
\begin{equation}
\begin{split}
 H = -2\sum_{j=1}^N \Big[\frac{1}{2}(S_j^+ S_{j+1}^- + S_j^- S_{j+1}^+) 
 + S_j^z S_{j+1}^z
 - S_j^z \Big] + \frac{3}{2}N, 
\end{split}
\end{equation}
which is the Hamiltonian of the XXX spin chain in a magnetic field. 
As was discussed in~\cite{bib:KMT99}, the third component of the total spin commutes with the Hamiltonian of the XXX chain, therefore the Bethe equations remain the same as those for the XXX chain without a magnetic field \eqref{be}. 

Next, we check the commutation relations with the fermion-number operator \eqref{commutation}. This is rather simple, since the fermion number operator just counts the number of down spins in the spin chain. Taking into account that the supercharge $Q$ removes a down spin, we have 
\begin{equation}
 [F,\,Q_j] 
  \begin{pmatrix}
   |\dots\,\underset{j-\frac{1}{2}}{\uparrow} \underset{j+\frac{1}{2}}{\downarrow}\,\dots\rangle \\
   |\dots\,\underset{j-\frac{1}{2}}{\downarrow} \underset{j+\frac{1}{2}}{\uparrow}\,\dots\rangle
  \end{pmatrix}
  = -Q_j 
  \begin{pmatrix}
   |\dots\,\underset{j-\frac{1}{2}}{\uparrow} \underset{j+\frac{1}{2}}{\downarrow}\,\dots\rangle \\
   |\dots\,\underset{j-\frac{1}{2}}{\downarrow} \underset{j+\frac{1}{2}}{\uparrow}\,\dots\rangle
  \end{pmatrix}. 
\end{equation}
On the other hand, $Q^{\dag}$ adds a down spin, that is, 
\begin{equation}
 [F,\,Q_j^{\dag}] 
  \begin{pmatrix}
   |\dots\,\underset{j}{\uparrow}\,\dots\rangle \\
   |\dots\,\underset{j}{\uparrow}\,\dots\rangle
  \end{pmatrix}
  = Q_j^{\dag} 
  \begin{pmatrix}
   |\dots\,\underset{j}{\uparrow}\,\dots\rangle \\
   |\dots\,\underset{j}{\uparrow}\,\dots\rangle
  \end{pmatrix}. 
\end{equation}
The topological charge of this model is calculated by considering the following linear combinations of supercharges: 
\begin{equation}
 Q^{(1)} = \frac{1}{2}(Q + Q^{\dag}), 
  \qquad
  Q^{(2)} = \frac{1}{2i}(Q - Q^{\dag}),  
\end{equation}
and subsequently, 
\begin{equation}
  q^{(1)} = \frac{1}{2}(q + q^{\dag}), 
  \qquad
  q^{(2)} = \frac{1}{2i}(q - q^{\dag}). 
\end{equation}
The operators $Q^{(1)}$ and $Q^{(2)}$ are not nilpotent. Instead, they satisfy 
\begin{equation}
 \langle \bm{\Psi}| (Q^{(1)})^2 |\bm{\Psi} \rangle = \langle \bm{\Psi}| (Q^{(2)})^2 |\bm{\Psi} \rangle = H, 
\end{equation}
while they anti-commute with $(-1)^F$: 
\begin{equation}
 \{(-1)^F,\, Q^{(1)}\} = \{(-1)^F,\, Q^{(2)}\} = 0. 
\end{equation}
Since the anti-commutation of the supercharges $Q^{(1)}$ and $Q^{(2)}$ gives the topological charge~\cite{bib:WO78}, we have 
\begin{equation}
 \{Q^{(1)},\, Q^{(2)}\} = \frac{1}{2i}(Q^2 - (Q^{\dag})^2) = 0, 
\end{equation}
which implies that the topological charge of the XXX model is zero.

\begin{figure}
 \begin{center}
  \includegraphics[scale=0.5]{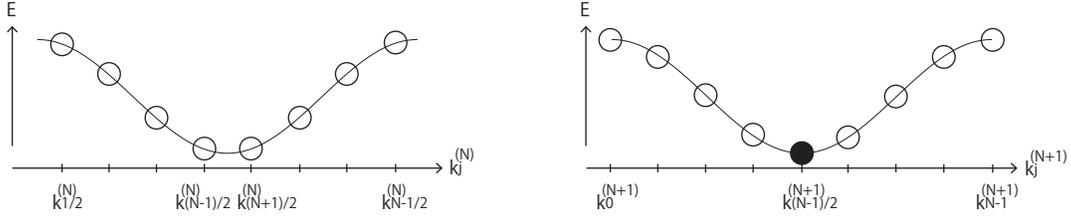}
  \caption{The discretized momenta of the XXX spin chain. The left figure is the dispersion relation for the even-length $N$ chain, while the right is that for the odd-length $(N+1)$ chain. The number of possible configurations of $n$ particle is the same in both chain, if we ignore the always-occupied quantum number denoted by a black dot. } \label{fig:quantized_momentum}
 \end{center}
\end{figure}

Finally, we comment on the actions of supercharges in the momentum space. As is well-known, the dispersion relation of the XXX spin chain is given by the trigonometric function: 
\begin{equation}
 E \sim \sum_{j=1}^n \cos k_j, 
\end{equation}
where the momentum is connected to the rapidity given as a Bethe root via 
\begin{equation} \label{momentum_rapidity}
 k_j = \frac{1}{i} \ln\frac{\sinh(\frac{i\gamma}{2} - \lambda_j)}{\sinh(\frac{i\gamma}{2} + \lambda_j)}. 
\end{equation}
On a finite length-$N$ chain, the momentum is quantized and allowed to take only discrete values: 
%\begin{equation} \label{quantized_momentum}
% k_j = \frac{2\pi}{N}j. 
%\end{equation}
\begin{equation} \label{quantized_momentum}
\begin{split}
 &j = \tfrac{1}{2}, \tfrac{3}{2}, \dots, \tfrac{N-1}{2}, \dots, N-\tfrac{1}{2}, \quad \text{for even $N$}, \\
 &j = 0, 1, \dots, \tfrac{N-1}{2}, \dots, N, \quad \text{for odd $N$}. 
\end{split}
\end{equation}
Notice that, in the XX ($\gamma \to \pi/2$) limit, the momenta coincide with those for the free fermions' $k_j = 2\pi j/N$. 
The vacancies of quantum numbers are denoted by white dots in Figure~\ref{fig:quantized_momentum}, whereas the quantum number always occupied is indicated by the black dot. 
From the relation between momentum and rapidity \eqref{momentum_rapidity}, the existence of rapidity $\lambda_{n+1} = i\pi$ indicates that there is a zero-momentum particle. This is compatible with the evidence in supersymmetry, where non-zero energy particles always make a superpartners. 
The Bethe ansatz analysis \eqref{be_matsui} implies that there exists a superpartner state for any set of Bethe roots of the length-$N$ chain. Therefore, there must be the same number of states in the length-$(N+1)$ chain if the zero-momentum quantum number is always occupied. This can be easily checked, since adding a site corresponds to adding a quantum number in the momentum space with one always-occupied quantum number, which results in the same degree of freedom before adding a site. 

\section{Supersymmetry in the spinon basis}
So far, we worked on the magnon basis, in which a down spin is regarded as a particle. In the magnon basis, a $n$-particle state is given by a superposition of the following-type state: 
\begin{equation}
 |\Psi(k_1,\dots,k_n)\rangle = \psi^{\dag}(k_1) \cdots \psi^{\dag}(k_n) |0\rangle, 
\end{equation}
where $|0\rangle$ is the Fock vacuum. $|\Psi\rangle$ is the $n$-particle state defined in the momentum space, and mapped to the rapidity state by the transformation \eqref{momentum_rapidity}: 
\begin{equation}
 |\Psi(\lambda_1,\dots,\lambda_n)\rangle = \psi^{\dag}(\lambda_1) \cdots \psi^{\dag}(\lambda_n) |0\rangle. 
\end{equation}
The rapidities $\lambda_j$ are given as Bethe roots, and thus the Bethe-ansatz description also follows the magnon basis. 

Let us write the action of supercharge in the momentum space. We consider the action of $Q$ on the state $|\Psi\rangle$. From the definition \eqref{supercharge_matsui}, we have 
\begin{align}
 Q |\Psi\rangle &= \sum_{j=1}^N 
 \begin{pmatrix}
  R_j \psi_{j-\frac{1}{2}} & 0 \\ 0 & R_j \psi_{j+\frac{1}{2}}
 \end{pmatrix}
  \psi^{\dag}(k_1) \cdots \psi^{\dag}(k_n) |0\rangle \\
 &=\sum_{k} \sum_{j=1}^N 
 \begin{pmatrix}
  R_j e^{-ik(j-\frac{1}{2})\frac{\pi}{N}} \psi(k) & 0 \\ 0 & R_j e^{-ik(j+\frac{1}{2})\frac{\pi}{N}} \psi(k)
 \end{pmatrix} 
 \psi^{\dag}(k_1) \cdots \psi^{\dag}(k_n) |0\rangle \\
 &= \psi(0) \psi^{\dag}(k_1) \cdots \psi^{\dag}(k_n) |0\rangle. \label{Q_action}
\end{align}
In the second line, we applied the Fourier transformation in order to write a supercharge in the momentum space. Then first summing up in the coordinate $j$, we obtained only the non-oscillating term in the third line. Notice that both of the non-zero elements acts as $\psi(0)$. The action of the supercharge \eqref{Q_action} that removes only a zero-momentum particle is quite compatible with the Bethe-ansatz analysis that tells us, by the action of $Q$, a magnon with $\lambda_{n+1} = i\pi$ {\it i.e.} zero momentum is removed. 
Note that $k$ takes only discrete values \eqref{quantized_momentum}. Thus, only the odd-length chain admits $k=0$. Thus, our discussion holds only on the odd-length chain for $Q$ and on the even-length chain for $Q^{\dag}$. This restriction was also pointed out in the previous work for the anisotropic case~\cite{bib:YF04}. 

Although we have been discussed the supercharges on the magnon basis so far, eigenstates of the spin chain also admits the spinon description, which is useful in comparison with the effective field theory. The emergence of a spinon in the spin-$1/2$ chain corresponds to the presence of a hole in the distribution of real roots. The spinon carries the rapidity of this hole. Therefore, in order to know how our supercharges behave in the spinon basis, we only need to check the difference of the number of holes before and after they act. 
The counting of admittable roots and holes are achieved by taking the logarithm of the Bethe equations~\cite{bib:FT81}. The logarithm of the Bethe equations \eqref{be} are obtained as 
\begin{equation} \label{be_log}
 N \ln \frac{\sinh(\lambda_j + \frac{i\gamma}{2})}{\sinh(\lambda_j - \frac{i\gamma}{2})}
  =
  \sum_{k=1 \atop k\neq j}^n \frac{\sinh(\lambda_j - \lambda_k + i\gamma)}{\sinh(\lambda_j - \lambda_k - i\gamma)} + 2\pi i M_j. 
\end{equation}
The quantum number $M_j$ is a half-integer for even $N$, while an integer for odd $N$. 
%Let us remark that one of the rapidity $\lambda_j$ is $i\pi$, otherwise the Bethe state is killed by $Q$. 
The roots may take complex numbers forming string solutions. If there are string-type roots, we need to take a product of those roots before converted into the logarithmic form. Then the equation \eqref{be_log} is given for string centers: 
\begin{align}
 &N \Phi_{r,1}(\lambda_j^{(r)}) = \sum_{r'} \sum_{k=1}^{\nu_{r'}} \Phi_{r,r'}(\lambda_j^{(r)} - \lambda_k^{(r')}) + 2\pi M_j^{(r)}, \\
 &\Phi_{r,r'}(x) = 2\sum_{k \in \{|n-m|, |n-m|+2, \dots, n+m-2\}}
 \left(\arctan\frac{2x}{k} + \arctan\frac{2x}{k+2}\right). 
\end{align} 
The string center of an $r$-string is denoted by $\lambda_j^{(r)}$. $\nu_{r'}$ is the number of $r'$-strings. The maximum value of the quantum number $M_{\rm max}$ is obtained from \eqref{be_log} by taking $\lambda_j \to \infty$ limit: 
\begin{equation} \label{max_qnumber}
 M^{(r)}_{\rm max} = \frac{1}{2\pi} 
  \left(N \Phi_{r,1}(\infty) - \sum_{r'}\nu_{r'} \Phi_{r,r'}(\infty)\right) - \frac{1}{2} 
\end{equation}
for even $N$. The odd $N$ case is similarly discussed. 
Thus, the admitted number of $r$-string roots is given by $2M_{\rm max}^{(r)} + 1$. The number of holes are obtained as the number of unoccupied quantum numbers. 

Now let us check how the number of holes in the distribution of real roots changes under the action of supercharges. For the the even $N$-length chain with $\nu_1=n$ roots, the calculation from \eqref{max_qnumber} leads to $(N-2n)$ holes. On the other hand, taking into account that, in the case of one less site with one less root $\lambda_{n} = i\pi$, the rest of roots remains the same, we have $(N-2n+1)$ holes. If we starts from the odd $(N+1)$-length chain with $\nu_1 = (n+1)$ roots, the number of holes are calculated as $N-2n-1$, while in the $N$-length chain with $\nu_1 = n$ roots, there are $(N-2n)$ holes. Thus, we obtain that the action of $Q$ adds a zero-momentum spinon, while the action of $Q^{\dag}$ removes it. For the cases with longer strings, the supercharges act in the same way. 
Since a spinon carries either spin-$1/2$ or spin-$(-1/2)$, it is interesting to know which kind of spinon is added or removed through the supercharge. We write the number of spinons with spin-$1/2$ by $n_+$ and those with spin-$(-1/2)$ by $n_-$. The total number of spinons $n_+ + n_-$ is given by the number of holes. Whereas, $n_+ - n_-$ gives the twice of the $z$-component of the total spin $2S^z$. Some examples of $(n_+,n_-)$ are calculated in Table~\ref{tab:spinon}. 
\begin{table}
 \begin{center}
 \begin{tabular}{cccc}
  \hline\hline
  & roots & $2S^z$ & $(n_+,n_-)$ \\
  \hline
  even $N$ & $\nu_1 = n$ & $N/2-n$ & $(N-2n,0)$ \\
  & $\nu_1=n-2,\, \nu_2=1$ & $N/2-n$ & $(N-2n+1,1)$ \\
  \hline
  odd $N+1$ & $\nu_1 = n+1$ & $(N-1)/2-n$ & $(N-2n-1,0)$ \\
  & $\nu_1=n-1,\, \nu_2=1$ & $(N-1)/2-n$ & $(N-2n,1)$ \\
  \hline\hline
 \end{tabular}
  \caption{The number of spinons for even $N$ and odd $N+1$. If a site is added together with a root, then a $1/2$-spinon is removed. } \label{tab:spinon}
 \end{center}
\end{table}
The supercharge $Q$ acts in such a way to add a spin-$1/2$ spinon, while $Q^{\dag}$ removes a spin-$1/2$ spinon. Therefore, if we write the spin-$\sigma/2$ spinon creation operator by $A^{\sigma}(\lambda)$ and the annihilation operator by $\widehat{A}^{\sigma}(\lambda)$, the supercharges act on the $n_s$-spinon state as 
\begin{align}
 &Q A^{\sigma_1}(\lambda_1) \cdots A^{\sigma_{n_s}}(\lambda_{n_s})|0\rangle 
  = A^{+}(i\pi) A^{\sigma_1}(\lambda_1) \cdots A^{\sigma_{n_s}}(\lambda_{n_s})|0\rangle, \label{spinon1} \\
% = \sum_{j=1}^n (-1)^{F_1}\psi^{\dag}(\theta_1) \cdots (-1)^{F_{j-1}}\psi^{\dag}(\theta_{j-1}) \psi^{\dag}(i\gamma)\psi^{\dag}(\theta_j) \dots \psi^{\dag}(\theta_n) |0\rangle, 
 &Q^{\dag} A^{\sigma_1}(\lambda_1) \cdots A^{\sigma_{n_s}}(\lambda_{n_s})|0\rangle 
 = \widehat{A}^{+}(i\pi) A^{\sigma_1}(\lambda_1) \cdots A^{\sigma_{n_s}}(\lambda_{n_s})|0\rangle. \label{spinon2}
% &= \frac{1}{n}\sum_{j=1}^n (-1)^{F_1}\psi^{\dag}(\theta_1) \cdots (-1)^{F_{j-1}}\psi^{\dag}(\theta_{j-1}) \psi(i\gamma)\psi^{\dag}(\theta_j) \dots \psi^{\dag}(\theta_n) |0\rangle \\
% &- (-1)^l \psi^{\dag}(\theta_1) \dots \psi^{\dag}(\theta_{l-1}) \psi^{\dag}(\theta_{l+1}) \dots \psi^{\dag}(\theta_n) |0\rangle 
\end{align} 
%That is, if we write the asymptotic in-state $(\theta_1 > \theta_2 > \dots \theta_n)$ as 
%\begin{equation}
% A(\theta_1) \cdots A(\theta_n) |0\rangle 
%  = |A(\theta_1) \cdots A(\theta_n) \rangle, 
%\end{equation}
%the supercharges acts as 
%\begin{align}
% &Q |A(\theta_1) \cdots A(\theta_n) \rangle 
%  = |A(\theta_1) \cdots A(\theta_n) A(i\gamma) \rangle, \\
% &Q^{\dag} |A(\theta_1) \cdots A(\theta_n) A(i\gamma) \rangle 
%  = |A(\theta_1) \cdots A(\theta_n) \rangle. 
%\end{align}
%Similarly for the out-state ($\theta_1 < \dots < \theta_n$), we have 
%\begin{align}
% &Q |A(\theta_1) \cdots A(\theta_n) \rangle 
%  = |A(i\gamma) A(\theta_1) \cdots A(\theta_n) \rangle, \\
% &Q^{\dag} |A(i\gamma) A(\theta_1) \cdots A(\theta_n) \rangle 
%  = |A(\theta_1) \cdots A(\theta_n) \rangle.  
%\end{align}
%The newly added or being-removed spinon $A(i\gamma)$ must be properly located in the above way, since it stays on the same position with respect to time development due to zero momentum. 
The second equation vanishes unless $\lambda_j \neq i\pi$ for any $j$. The commutation relations of $A^{\sigma}$'s are derived through the mode expansions as in~\cite{bib:BLS94}. It is also an interesting problem, which we do not discuss in this paper, to compare the spinon basis with the soliton basis~\cite{bib:L08, bib:L08}, in which scattering amplitudes are given by the Boltzmann weights of the six-vertex model~\cite{bib:ZZ79}.  
%\begin{equation}
% A^{\sigma}(\lambda') A^{\sigma'}(\lambda) 
%  = R^{\sigma\sigma'}_{\sigma\sigma'}(\lambda-\lambda') A^{\sigma}(\lambda) A^{\sigma'}(\lambda')
%  + R^{\sigma\sigma'}_{\sigma'\sigma}(\lambda-\lambda') A^{\sigma'}(\lambda) A^{\sigma}(\lambda'), 
%\end{equation}
%where the $R$ is the $R$-matrix of the six-vertex model: 
%\begin{equation}
%\begin{split}
% &R^{++}_{++}(\lambda) = R^{--}_{--}(\lambda) = 1, \\
%&R^{+-}_{+-}(\lambda) = R^{-+}_{-+}(\lambda) = \frac{\sinh\lambda}{\sinh(\lambda + i\gamma)}, \\
%&R^{+-}_{-+}(\lambda) = R^{-+}_{+-}(\lambda) = \frac{i\sin\gamma}{\sinh(\lambda + i\gamma)}. 
%\end{split}
%\end{equation}

\begin{figure}
 \begin{center}
  \includegraphics[scale=0.7]{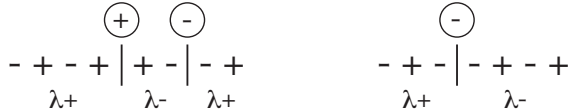}
  \caption{The domain walls and domains for an even length chain (left) and an odd length chain (right). In the even length chain, both edges belong to the same domain. On the other hand, in the odd length chain, both edges belong to the different domains. Thus, the even number of spinons are always obtained in the even length chain, while only the odd number of spinons are obtained in the odd length chain. Spinons are located at the positions of domain walls. The sign of spinon-spin is determined by the signs of both sides of walls. } \label{fig:domain_walls}
 \end{center}
\end{figure}

Although the supercharges act to change the number of spinons by one, we would rather interpret them as the actions on domain walls. 
%it is helpful for comparison with the conventional actions to give an interpretation of supercharges without changing the number of particles. 
In order to regard a spinon as a quantum kink, we first rewrite states of the spin chain in terms of domains and domain walls~\cite{bib:I93}. Let us consider the classical anti-ferromagnetic state $(-\,+\,-\,+\,-\,+\,\dots)$. We call the domain characterized by this configuration the $\lambda_+$-domain. There is another anti-ferromagnetic state which flips all spins of the $\lambda_+$-domain $(+\,-\,+\,-\,+\,-\,\dots)$, which we call the $\lambda_-$-domain. 
%The notion of domains and domain walls in the spin chain is often used to discuss the level-$k$ WZW model, the effective field theory of the spin-$k/2$ spin chain, whose local states are characterized by the RSOS indices which takes $(k+1)$ values, besides the spinon degree of freedom. In the spin-$1/2$ case, only the spinon degree of freedom survives, however, spinon states are described as domains and domain walls, like the $2$-value RSOS model. 
Now we set the domain walls between the two different domains. Since there are only two types of domains, only two kinds of domain walls are obtained. Two domains must alternatively appear in such a way that the wall $(\lambda_+,\lambda_-)$ and $(\lambda_-,\lambda_+)$ are alternatively located. On both sides with the walls, there are always the same direction of two spins. These spins form a kink, which is to be interpreted as a spinon in the quantum system. Thus, a spinon always characterized by the domains on its both sides. Besides, a spinon carries spin-$(\pm 1/2)$~\cite{bib:FT81}. The sign depends on whether the spinon is generated from two up spins or two down spins. If it is formed by two up spins, the spinon carries spin-$1/2$, while if it is from two down spins, the spinon carries spin-$(-1/2)$. 
Thus, a spinon is an excitation quasi-particle on the anti-ferromagnetic ground state. Due to the spin-reversal symmetry, both $\lambda_+$ and $\lambda_-$ can be the ground-state domain. Here we take $\lambda_+$ as the physical domain. Since the even-length chain and the odd-length chain behaves in different ways, we consider these two cases separately. 
In the even length chain, the ground state consists only of the $\lambda_+$-domain. In an excitation state, only an even number of spinons are allowed to exist. Therefore, both edges of the even length chain belong to the same domain. On the other hand, in the odd length chain, the existence only of an odd number of spinons are permitted. This implies that there is no true ground state (Figure~\ref{fig:domain_walls}). For this reason, the odd length chain has an edge of the $\lambda_+$-domain, while the other with the $\lambda_-$-domain. 
The emergence or disappearance of a spinon naturally explains this modification of the edge domains. That is, a zero-momentum spinon is regarded to be an object to exchange the domains on its both side, rather than a particle which itself carries a domain wall. 

%In the conventional notation, the action of the supercharges on the $n$-particle state of a kink type are known to be written as~\cite{} 
%\begin{align}
% &Q^{(1)} |A(\theta_1) \cdots A(\theta_n) \rangle
%  = \sqrt{m} \sum_{j=1}^n |(A(-1)^F)(\theta_1) \cdots (Aq^{(1)})(\theta_j) \cdots A(\theta_n) \rangle, \label{spinon_action1}\\
% &Q^{(2)} |A(\theta_1) \cdots A(\theta_n) \rangle
%  = \sqrt{m} \sum_{j=1}^n |(A(-1)^F)(\theta_1) \cdots (Aq^{(2)})(\theta_j) \cdots A(\theta_n) \rangle, \label{spinon_action2}
%\end{align}
%where $m$ is a particle mass. The supercharges $Q^{(1)}$ and $Q^{(2)}$ are related to our supercharges through 

\section{Conclusions}
In this paper, the supersymmetry was found in the spin-$1/2$ isotropic Heisenberg chain by constructing the supercharges in terms of the Jordan-Wigner fermions. The actions of the supercharges were dynamical on the coordinate basis, however, we introduced the new interpretation of these actions in the spinon basis, on which creation or annihilation of a zero-momentum spinon is regarded as exchange of the edge domains. 

Although the superalgebraic structure of the dynamical operators was already investigated at the $\gamma = \pi/3$ point in~\cite{bib:YF04}, it is important to know how they work on the spinon basis for applying the method to continuum systems such as the $\mathcal{N} = 4$ SYM of the higher-loop level, since in the quantum field theory, a supercharge is defined in the particle basis. 
However, our supercharges are well-defined only on an odd length chain for the site-decreasing operator, while on an even length chain for the site-increasing operator. Actually, if it is forbidden to have zero-momentum spinon, there is no superpartner of an arbitrary non-zero energy state. Therefore, it is an interesting problem whether the supersymmetry in nature is defined only on the odd length chain or there is some way to define it also on the even length chain. 

The motivation of this work is a discovery of the Bethe root which corresponds to the zero-momentum in the same Bethe equations as those for the one-more site without the zero-momentum roots. Another supersymmetric point, {\it i.e.} the Razumov-Stroganov point can be also found in a similar way, which motivates us to state that there would be the other supersymmetric points. Starting from the Bethe equations for the length-$N$ chain with $n$ Bethe roots \eqref{be}, we may deform it as 
\begin{equation}
 \left(\frac{\sinh(\lambda_j + \frac{i\gamma}{2})}{\sinh(\lambda_j - \frac{i\gamma}{2})}\right)^N
  \frac{\sinh(\lambda_j + is\gamma)}{\sinh(\lambda_j - is\gamma)}
  = 
  \prod_{k=1 \atop k \neq j}^n 
  \frac{\sinh(\lambda_j - \lambda_k + i\gamma)}{\sinh(\lambda_j - \lambda_k - i\gamma)}
  \frac{\sinh(\lambda_j + is\gamma)}{\sinh(\lambda_j - is\gamma)}, 
\end{equation}
which is the Bethe equations for the length-$(N+1)$ with $(n+1)$ roots with spin-$s$ at the $(N+1)$th site. These Bethe equations posses the solution $\lambda_{n+1} = i\pi$ for the anisotropy $\gamma = \pm\pi/(s-1)$, which is expected to be another supersymmetric point. Thus, we expect that the supersymmetry structure would be found in spin chains with magnetic impurities.

\section*{Acknowledgements}
The author would like to thank M. Staudacher, V. Mitev, and H. Katsura for helpful discussion. This work is supported by Grant-in-Aid for Young Scientists (B) of the grant number 15K20939.

%Bibliography
\bibliographystyle{unsrt}
\bibliography{reference}

\end{document}